\documentclass[prl, twocolumn, showpacs, english, preprintnumbers, amsmath, amssymb, superscriptaddress, aps]{revtex4-1}
\usepackage{latexsym}
\usepackage{epsfig,psfrag}
\usepackage{dcolumn}
\usepackage{bm}
\usepackage{graphicx}
\usepackage{color}
\usepackage{esint}
\usepackage{babel}
\usepackage{amsfonts}
\usepackage{enumerate}
\begin{document}
\title{Non-Fermi liquid manifold in a Majorana device} 

\author{Erik Eriksson,$^1$ Christophe Mora,$^2$ Alex Zazunov,$^1$ 
and Reinhold Egger}
\affiliation{Institut f\"ur Theoretische Physik,
 Heinrich-Heine-Universit\"at, D-40225 D\"usseldorf, Germany\\
$^2$~Laboratoire Pierre Aigrain, {\'E}cole Normale Sup{\'e}rieure,
Universit{\'e} Paris 7 Diderot, CNRS;\\ 24 rue Lhomond, F-75005 Paris, France}
\date{\today}

\begin{abstract}
We propose and study a setup realizing a stable manifold of non-Fermi 
liquid states.  The device consists of a mesoscopic 
superconducting island hosting $N\ge 3$ Majorana bound states 
tunnel-coupled to normal leads, with a Josephson contact to a
bulk superconductor.  We find a nontrivial interplay between
multi-channel Kondo and resonant Andreev reflection processes, 
which results in the fixed point manifold.  
The scaling dimension of the leading irrelevant perturbation changes 
continuously within the manifold and determines the 
power-law scaling of the temperature dependent conductance. 
\end{abstract}

\maketitle

\begin{figure}
\centering
\includegraphics[width=6cm]{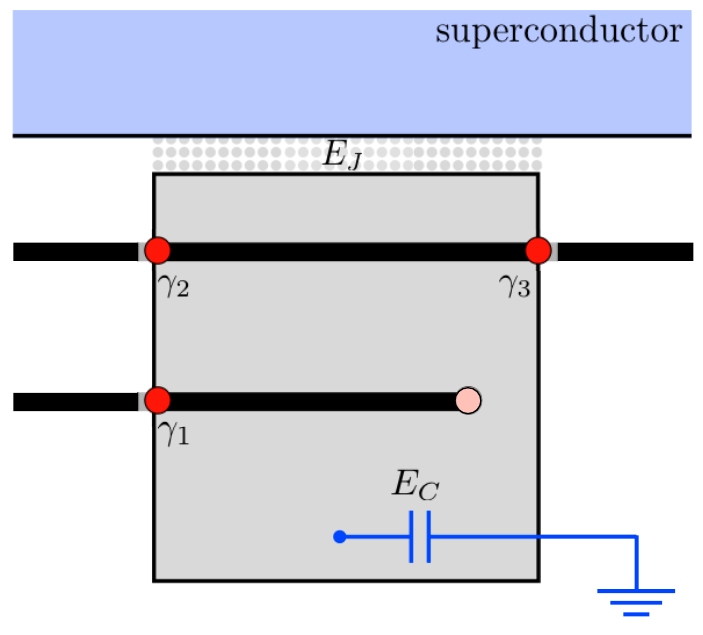}
\caption{\label{fig1} (Color online) 
Schematic device setup: Several one-dimensional (1D) 
nanowires with strong spin-orbit coupling are deposited on
a floating superconducting island with charging energy $E_C$.
Choosing appropriate system parameters, 
see Refs.~\cite{mbsrev1,mbsrev2,mbsrev3} for a thorough discussion, 
each nanowire hosts two spatially separated Majorana bound states.
The overhanging parts of the wire act as normal-conducting leads, 
where only effectively spinless 1D fermions, $\Psi_j(x)\sim \eta_j+i\rho_j$, 
couple to the island.  Of the $N_{\rm tot}$ Majorana states on the island,
$N$ are connected to leads (here $N=3$), where the other 
$N_{\rm tot}-N$ Majoranas have no effect on the physics described here.
The island also couples to a bulk superconductor through the 
Josephson energy $E_J$. 
}
\end{figure}

\textit{Introduction.---}Nanoscale devices hosting Majorana bound states 
are expected to display spectacular non-local quantum correlations and
long-range entanglement \cite{hasan,mbsrev1,mbsrev2,mbsrev3}. 
Experimental reports of Majoranas fermions 
\cite{exp1,exp2,exp3,exp4,exp5,exp6} have so far focused on
effectively noninteracting systems, where local resonant Andreev 
reflection (RAR) physics dominates the transport 
characteristics \cite{mbsrev1}.  Since interactions tend to suppress RAR, 
several interesting non-local phenomena have been predicted for 
interacting Majorana devices, such as electron
teleportation \cite{fu2010,zazu2011,roland}, 
interaction-induced unstable fixed points \cite{altland1,beri2}, or 
the topological Kondo effect \cite{beri1}, where strong charging effects 
cause a multi-channel Kondo state. As a general rule, such states display
non-Fermi liquid (NFL) behavior
\cite{blandin,wiegmann,andrei,tsvelik2,hewson,gogolinbook}.
The Kondo and RAR states, resp., constitute mutually exclusive phases
in all settings studied up to now \cite{beri1,altland1,beri2,lutchyn}. 
In this paper, we predict that a nontrivial coexistence of Kondo and 
RAR physics takes place in the device shown in Fig.~\ref{fig1}, 
where a mesoscopic superconducting island 
is Josephson coupled to a conventional bulk  superconductor 
and hosts $N\ge 3$ Majoranas weakly contacted by normal leads.  
In principle, all ingredients are experimentally available 
\cite{exp1,exp2,exp3,exp4,exp5,exp6}.  We find that the Kondo-RAR interplay 
in such a device can result in a continuously 
tunable \textit{manifold} of NFL states.  Although similar physics was 
proposed before for conventional Kondo systems 
\cite{georges1,georges2,ye,garst,fiete2}, 
anisotropies destabilize the corresponding NFL fixed points and have 
prevented their experimental observation.
In our proposal, the stability of the NFL manifold 
is tied to the non-local Majorana representation of 
an effective ``quantum impurity'', where Kondo screening 
and RAR processes both originate from 
the tunnel coupling between Majoranas and lead electrons.  

Before entering a detailed discussion, we briefly summarize our
main results.  
The low-energy physics near the ground-state NFL manifold is governed
by a leading irrelevant perturbation of scaling dimension  
\begin{equation}\label{scaldim}
y ={\rm min}\left\{ 2, \frac12 \sum_{j=1}^N   
\left[ 1- \frac{2}{\pi} \sin^{-1}  \left(
\frac{\delta_j}{2(N-1)}\right) \right]^2\right\} ,
\end{equation}
where the $N$ dimensionless parameters $\delta_j= \sqrt{\Gamma_j/T_K}$
depend on the lead-to-Majorana hybridizations, $\Gamma_j$, and 
the Kondo temperature, $T_K$, the respective energy scales for RAR and 
Kondo physics.   The $(\delta_1,\ldots,\delta_N)$ domain with 
$y>1$ corresponds to the NFL manifold, which could be explored 
experimentally by varying the $\Gamma_j$ via gate voltages \cite{exp1}.  
The NFL character is manifest in the non-integer 
and continuously tunable scaling dimension in Eq.~(\ref{scaldim}).
Interestingly, a similar low-energy model has been obtained for 
the two-channel two-impurity Kondo model, despite of a different
physical origin, where scaling dimensions and finite-size spectra
were derived in Refs.~\cite{georges1,georges2}.
Our predictions can be observed in charge transport, 
since $y$ governs the power-law scaling of the temperature-dependent 
conductance tensor at $T\ll T_K$,
\begin{equation}  \label{condtens}
G_{jk}(T)= \frac{2e^2}{h} \left[ \delta_{jk} - A_{jk} 
\left(\frac{T}{T_K}\right)^{2(y-1)} +\cdots \right],
\end{equation}
with dimensionless numbers $A_{jk}(\delta_1,\ldots,\delta_N)$ 
of order unity.  Albeit Eq.~\eqref{condtens}  
coincides with the local RAR result \cite{mbsrev1} for $T=0$, 
it reflects entirely different physics. This difference is 
readily observable at finite $T$, 
where the \textit{non-local}\ conductances $G_{j\ne k}$ in Eq.~\eqref{condtens}
are finite, in marked contrast to the RAR case.

\textit{Device proposal.---}We consider the setup in Fig.~\ref{fig1}, where
a floating mesoscopic superconducting island, with charging energy
$E_C$, is in proximity to at least two nanowires with strong 
spin-orbit coupling, e.g., InSb or InAs.  The island's superconducting
phase, $\varphi$, is taken relative to a conventional bulk superconductor,
where the Josephson energy, $E_J$, denotes their coupling and we 
assume a large pairing gap such that quasiparticle poisoning is negligible.  
In the presence of a Zeeman field, Majorana bound states are induced
near each end of a superconducting nanowire part
\cite{exp1,exp2,exp3,exp4,exp5,exp6,mbsrev1,mbsrev2,mbsrev3}.
We study the case that $N\ge 3$ Majoranas, 
described by operators $\gamma_j=\gamma_j^\dagger$
with anticommutator algebra $\{ \gamma_j,\gamma_k \}=\delta_{jk}$,
are connected to normal leads.   We assume that different
Majoranas are well separated, i.e., direct tunnel couplings
can be neglected. Note that their distance may exceed the superconducting
coherence length since the phase dynamics of the 
Cooper pair condensate renders
transport intrinsically non-local in such a device \cite{fu2010}.  
The island Hamiltonian, $H_{\rm island} = 
E_C \left(Q-n_g\right)^2 -E_J \cos\varphi$,
then contains a charging and a Josephson energy contribution, respectively.
The total electron number on the island, $Q$, is due to
Cooper pairs and occupied Majorana states \cite{fu2010,zazu2011,roland}, 
and the backgate parameter $n_g$ has no effect in the regime studied below.
Using units with $\hbar=k_B=1$, the Hamiltonian, $H=H_0+H_t+H_{\rm island}$, 
also contains a lead part, $H_0= -i v_F \sum_{j} \int_{-\infty}^\infty dx \
\Psi^\dagger_j  \partial_x \Psi_j^{}$,
with Fermi velocity $v_F$.  In each lead, only an effectively 
spinless chiral 1D fermion, $\Psi_j(x)$, corresponding to the
overhanging wire parts in Fig.~\ref{fig1}, connects to the island by
tunneling via the Majorana fermion $\gamma_j.$ 
This is described by the tunneling Hamiltonian \cite{zazu2011}, 
$H_t=  \sum_{j=1}^N \ \lambda_j e^{-i\varphi/2} \Psi_j^\dagger (0) 
\gamma_j + {\rm h.c.}$, where the tunnel couplings $\lambda_j$ 
are chosen real positive and $x=0$ marks the contact.  
With hybridization parameters $\Gamma_j=2\pi \nu_0\lambda _j^2$,
the lead density of states $\nu_0=1/\pi v_F$, and the 
Josephson plasma frequency $\Omega=\sqrt{8 E_C E_J}$,
the regime of interest is ${\rm max}(\Gamma_j) \ll \Omega \alt E_J$.  
In presently studied experimental devices \cite{exp1,exp7}, 
both the pairing gap and the charging energy of the island are of
the order of a few~meV.  Choosing also the value of $E_J$ --
which mainly depends on the interface to the bulk superconductor --
within the meV regime, and noting that the hybridizations are also
gate-tunable with $\Gamma_j\approx 0.01\ldots 1$~meV \cite{exp1}, 
the implementation of our proposal seems possible.  The observation of
the predicted phenomena also requires low temperatures, 
$T\ll T_K$, see below. 

\textit{Effective low-energy Hamiltonian.---}We next show that 
for ${\rm max}(\Gamma_j) \ll \Omega \alt E_J$, 
a simpler effective low-energy theory emerges.
In this regime, the phase $\varphi$ will mostly stay near 
the minima of the $-E_J\cos\varphi$ term in $H_{\rm island}$.
Phase slips due to tunneling between adjacent minima
are exponentially suppressed \cite{zaikin},
and it is justified to neglect them.  The phase dynamics then consists of fast
zero-point oscillations of frequency $\Omega$ around a given minimum. 
Since $\langle (\delta \varphi)^2 \rangle=\Omega/2E_J$, 
the oscillation amplitude remains small and we may 
integrate over the $\varphi$ fluctuations. 
The resulting effective low-energy Hamiltonian,
$H_{\rm eff} = H_0+ H_A+H_K$, is local on timescales above $\Omega^{-1}$.
Expressing the lead fermions by pairs of chiral Majorana fields, 
$\Psi^{}_j(x) =[\eta_j(x)+i\rho_j(x)]/\sqrt{2}$, we obtain 
\begin{eqnarray} 
\label{newtunnH}
&& H_0 = -\frac{iv_F}{2}\sum_{j=1}^N \int dx \left(\eta_j\partial_x\eta_j+
\rho_j\partial_x\rho_j\right),\\ 
\nonumber
&& H_A = \sqrt{2}i \sum_{j} \lambda_j \gamma_j \rho_j(0) , 
\quad H_K = \sum_{j<k} J_{jk} \gamma_j\gamma_k \eta_k(0) \eta_j(0).
\end{eqnarray}
The positive ``exchange couplings'', $J_{jk} =\lambda_j \lambda_k/4E_J$, are
controlled by $E_J$.  Although  $E_C$  does not appear in $H_{\rm eff}$,
it enters the bandwidth given by the plasma frequency $\Omega$.  
In Eq.~\eqref{newtunnH}, $H_A$ couples only to $\rho_j$ and
describes RAR \cite{mbsrev1}, while $H_K$ only involves 
the $\eta_j$ Majoranas and describes exchange processes between 
lead electrons and the components $\gamma_j\gamma_k$ of 
the ``impurity spin''.  On top of terms 
$\sim \Psi_j^\dagger(0)\Psi^{}_k(0)$, which also appear in the 
topological Kondo model of Ref.~\cite{beri1}, $H_K$ contains crossed 
Andreev reflection contributions, e.g., terms 
$\sim \Psi_j^\dagger(0)\Psi_k^\dagger(0)$, where a Cooper pair
splits into two electrons in separate leads.
Due to the phase coherence in the superconductor,
which is behind the $e^{\mp i \varphi/2}$ phase factors in $H_t$,
both types of exchange processes enter $H_K$ with equal weight. 
Without the $H_A$ term, $H_{\rm eff}$ is 
mathematically identical to the SO$_1(N)$ Kondo model recently proposed 
for crossed Ising chains, which hosts a NFL Kondo fixed 
point \cite{tsvelik,crampe} and, for $N=3$, is equivalent 
to the conventional two-channel Kondo model because of the group relation
 SO$_1(3) \sim$~SU$_2(2)$.

\textit{Renormalization group analysis.---}By employing
standard energy-shell integration \cite{hewson}, 
we find the one-loop renormalization group (RG) equations 
\begin{equation}\label{rgeq}
\frac{d\Gamma_j}{dl}= \Gamma_j, \quad 
\frac{dJ_{j\ne k}}{dl} = 2\nu_0 \sum_{m\ne (j,k)} 
\frac{J_{jm} J_{mk}}{1+\Gamma_m/\Omega}.
\end{equation}
The running couplings $\Gamma_j(l)$ thus approach the strong coupling
limit according to the standard RAR equations \cite{mbsrev1},
while the RG flow of the exchange couplings is coupled to 
the $\Gamma_j$. Similar to what happens in the pure Kondo 
case \cite{tsvelik}, Eq.~\eqref{rgeq} implies that 
anisotropies in the $J_{jk}$ are RG irrelevant, 
while the isotropic part is marginally relevant.  We thus 
write $J_{jk}=J(1-\delta_{jk})$, and neglect irrelevant deviations from 
isotropy from now on.  
We shall also assume $\Gamma_j=\Gamma$, but return
to the role of $\Gamma_j$ anisotropy later.  
To one-loop accuracy, we then obtain the estimate
$T_K \approx \Omega \exp\left(- \frac{E_J}{(N-2)\Gamma} \right)$ for 
 the Kondo temperature.  Moreover, 
Eq.~\eqref{rgeq} can now be solved analytically.  This solution
shows that both $\Gamma(l)$ and $J(l)$ flow towards
 strong coupling for $\Gamma<T_K$.  Especially for large $N$,
it is possible to satisfy this condition by choosing $\Omega\approx E_J$ and
not too small ratio $\Gamma/E_J$.  
In what follows, we focus on the regime $\Gamma<T_K$, and analyze the
physics at low temperatures, $T\ll T_K$.   For $\Gamma>T_K$, 
one instead arrives at the well known RAR picture \cite{mbsrev1}.
To estimate the Kondo scale for typical parameters, let us put, say, $N=6$,
$\Gamma= 0.2$~meV and $\Omega=E_J=2$~meV, where 
$T_K\approx 0.27~$meV, and $\Gamma<T_K$ is satisfied.
The low-temperature regime with $T\ll T_K$ is then also
accessible to experiments.

\begin{figure}
\centering
\includegraphics[width=7cm]{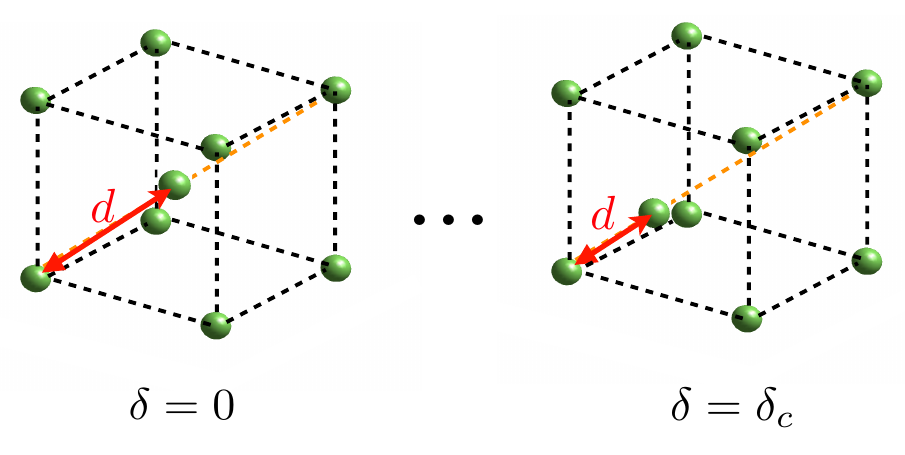}
\caption{\label{fig2} Lattice corresponding to the potential
minima of $V[\boldsymbol{\Phi}]$ for $N=3$, with $\delta=0$ (left)
and $\delta=\delta_c$ (right).  With increasing $\delta$, the center
of the lattice moves along the diagonal towards the corner point.
The line of fixed points (corresponding to the non-Fermi liquid manifold
for $\delta_1=\delta_2=\delta_3$) terminates at $\delta=\delta_c$.
}
\end{figure}

\textit{Quantum Brownian motion analogy.---}The low-temperature 
physics within the most interesting regime $\Gamma<T_K$ can be 
captured from an instructive analogy to quantum Brownian motion in a 
lattice-periodic potential.
To see this, we first bosonize the lead fermions in $H_{\rm eff}$ by writing
$\Psi_j(x) = \xi_K^{-1/2} \zeta_j 
e^{i\phi_j(x)}$ \cite{gogolinbook}
with boson fields $\phi_j(x)$, where $\xi_K=v_F/T_K$ sets the 
short-distance scale and additional Majorana fermions, $\zeta_j$, 
represent the Klein factors enforcing fermion 
anticommutators between different leads \cite{gogolinbook}.
 Following Refs.~\cite{altland1,beri2}, each ``true''
Majorana fermion, $\gamma_j$, is combined with the respective ``Klein'' 
Majorana,
 $\zeta_j$, to form an auxiliary fermion. The latter has conserved
occupation number and can be gauged away \cite{altland1}.
This yields a purely bosonic action, $S[\boldsymbol{\Phi}]=
\sum_{j} \int \frac{d\omega}{2\pi} |\omega|\left|\tilde\Phi_j( 
\omega)\right|^2+ \int d\tau\ V[\boldsymbol{\Phi}(\tau)]$, 
where $\boldsymbol{\Phi}=(\Phi_1,\ldots,\Phi_N)$ with
$\Phi_j\equiv\phi_j(x=0)$ and 
Fourier components $\tilde\Phi_j(\omega)$. The Gaussian part
describes dissipation by electron-hole pair excitations in the leads,
and the RAR-Kondo interplay is encoded by the ``pinning potential''
\begin{equation} \label{hefft}
V\left[\boldsymbol{\Phi}\right] = -\frac{\lambda}{\sqrt{\xi_K}} 
\sum_{j} \sin\Phi_j - \frac{J}{4\xi_K} \sum_{j\ne k} \cos \Phi_j 
\cos \Phi_k.
\end{equation}
We thus arrive at the quantum Brownian motion of a 
fictitious particle with coordinates
$\boldsymbol{\Phi}$ in the $N$-dimensional 
lattice corresponding to $V[\boldsymbol{\Phi}]$, see also
Refs.~\cite{YiKane,YiQBM}.  
Comparison to the $N=3$ field theory (see below) shows that, up to an 
overall prefactor, the renormalized couplings $\lambda$ and $J$ in 
Eq.~\eqref{hefft} are effectively replaced by 
$\sqrt{\xi_K\Gamma}$ and $4\xi_K\sqrt{T_K}$, respectively, 
when approaching the strong-coupling regime.  The relative importance
of the two terms in Eq.~\eqref{hefft} is thus governed by
$\delta=\sqrt{\Gamma/T_K}$.

In the ground state, $\boldsymbol{\Phi}$ is pinned to one of the 
minima of $V[\boldsymbol{\Phi}]$.  These minima occur for 
isotropic boson field configurations, $\Phi_j=\Phi_{\rm min}$,
with $\sin\left(\Phi_{\rm min}\right)=
\delta/[2(N-1)]$.  For $\delta=0$, the minima at $\Phi_{\rm min}=0$ and 
$\Phi_{\rm min}=\pi$ correspond
to the corner and center points, respectively, 
of a body centered hyper-cubic lattice. 
These points move in opposite directions when increasing $\delta$,
such that we have two interpenetrating cubic lattices. The closest
distance between corner and center points, see Fig.~\ref{fig2} for
an illustration, is given by $d=\sqrt{N}(\pi-2\Phi_{\rm min})$, 
while the distance between corners (or between centers) remains $d=2\pi$.
Perturbations around the ground state then come from instanton transitions
connecting different potential minima.  Following the arguments 
of Yi and Kane \cite{YiKane,YiQBM}, the scaling dimension $y$ of
the perturbation is directly related to the distance $d$ between the potential
minima, $y= d^2/(2\pi^2)$. 
For the leading (nearest-neighbor) term, we arrive
at Eq.~\eqref{scaldim} announced above.  
This  perturbation is RG irrelevant for $\delta<\delta_c$, with
\begin{equation}\label{deltacc}
\delta_c = 2(N-1) \sin\left[\frac{\pi}{2}\left(1-\sqrt{\frac{2}{N}}
 \right) \right]. 
\end{equation}
Since $y(\delta)$ is not an integer, all stable fixed points can be
classified as NFL states.  As a consequence, we obtain a 
stable line of NFL fixed points parametrized by 
$0\le \delta<\delta_c$.  
For $\delta>\delta_c$,  the perturbation becomes relevant and 
destabilizes the fixed point line.  Since this 
corresponds to $\Gamma>T_K$, we conclude that $\delta_c$ 
marks the phase transition to the RAR regime.

\textit{Strong coupling approach.---}It is reassuring
that the above results can be confirmed by an 
explicit strong-coupling solution for $N=3$, which we 
briefly sketch next.  Encoding the Majorana triplet 
 $\boldsymbol{\gamma}=(\gamma_1,\gamma_2,\gamma_3)$
by a spin-$1/2$ operator,
$\boldsymbol{S} = -(i/2) \boldsymbol{\gamma}\times \boldsymbol{\gamma}$,
plus another Majorana fermion, 
$b=-2i\gamma_1\gamma_2\gamma_3$ \cite{wilczek},  
the RAR term in Eq.~\eqref{newtunnH} reads
$H_A=  2\sqrt{2}i \lambda  b \boldsymbol{S}\cdot \boldsymbol{\rho}(0)$,
while the Kondo term becomes
$H_K = J \boldsymbol{S}\cdot \left[ -\frac{i}{2} \boldsymbol{\eta}(0)\times
\boldsymbol{\eta}(0) \right]$.
We now recall that without the RAR term, $H_{\rm eff}$ reduces to
the standard two-channel Kondo model, where the results of
Refs.~\cite{georges1,georges2,emery,georges3,coleman,sela1,sela2,lehur}
imply: (i) The $\boldsymbol{\eta}$ triplet of lead Majoranas 
obeys twisted boundary conditions, 
$\boldsymbol{\eta}(x)\to {\rm sgn}(x) \boldsymbol{\eta}(x)$.
The sign change when passing the impurity implies that
an incoming electron is effectively reflected as a hole with 
unit probability.  This resembles the RAR mechanism and rationalizes why
the $T=0$ conductance in Eq.~\eqref{condtens} coincides with the
RAR result.  (ii) Screening processes, entangling the impurity spin
with $\boldsymbol{\eta}$, are effectively described by writing 
$\boldsymbol{S} = i \sqrt{\xi_K}  a \boldsymbol{\eta}(0)$,
where $a$ is a new Majorana fermion capturing the remaining unscreened 
degree of freedom.
(iii) The leading irrelevant operator corresponds to 
$H'_K = 2\pi T_K \xi_K^{3/2} a \eta_1(0)\eta_2(0) \eta_3(0).$

Including now the RAR term, $\lambda\ne 0$, 
we combine the $a$ and $b$ Majoranas to
a conventional fermion, $d=(a+ib)/\sqrt{2}$.  Using (ii) and bosonizing
the lead fermions as above, the low-energy form of the RAR contribution is
$H'_A= (\sqrt{6} v_F \delta/\pi) [d^\dagger d-1/2] \partial_x \phi_0(0)$,
with $\phi_0=(\phi_1+\phi_2+\phi_3)/\sqrt{3}$. This expression is
reminiscent of the X-ray edge singularity problem \cite{gogolinbook}, 
suggesting that the marginal perturbation $H'_A$ can be
nonperturbatively included into $H_0$ by
a unitary transformation.  Indeed, with $U=e^{i (2\sqrt{6} \delta/\pi) 
(d^\dagger d-1/2) \phi_0(0)}$, this is the case, 
where $UH'_K U^\dagger$ generates eight different operators.
The smallest scaling dimension,
$y(\delta)=(3/2) \left[1-\delta/(2\pi)\right]^2$,
then identifies the leading irrelevant operator 
\cite{georges1,georges2}. This result is exact for $\delta\ll 1$, where
it matches Eq.~\eqref{scaldim}.
Stability requires $\delta<\delta_c= 2\pi (1-\sqrt{2/3})\simeq 1.153$,
in good agreement to the value 
predicted by Eq.~(\ref{deltacc}), $\delta_c\simeq  1.137$.

\textit{Discussion.---}So far we have studied the isotropic
setup with $\Gamma_j=\Gamma$.  While anisotropic deviations in 
the exchange couplings $J_{jk}$ are RG irrelevant,  
deviations in the $\Gamma_j$ convert the fixed point line 
into an $N$-dimensional manifold parametrized by the
$\delta_j=\sqrt{\Gamma_j/T_K}$.  In the quantum Brownian motion
approach, the $\boldsymbol{\Phi}$ potential minima then  
move away from isotropic configurations, and
$y=y(\delta_1,\ldots,\delta_N)$ in Eq.~(\ref{scaldim}) has been 
obtained by computing
the distance $d$ between nearest neighbor minima.
The resulting NFL can be probed in charge transport experiments.
The conductance tensor is defined by
$G_{jk}(T)=-e \partial I_j/\partial \mu_k$, where 
the $j$th lead has chemical potential $\mu_j$, and the charge currents $I_j$
are oriented towards the island.  
Closely following the technical steps detailed in Ref.~\cite{njpZ},
their steady-state expectation values can be obtained
from a Keldysh functional integral, since the fixed point theory
is represented by a Gaussian action for the \textit{dual}\ boson fields.
Perturbation theory in the leading irrevelant perturbation,
of scaling dimension (\ref{scaldim}), then
determines the linear conductance tensor
for $T\ll T_K$ as stated in Eq.~\eqref{condtens}.
For $\delta_j=\delta$, all matrix elements $A_{jk}(\delta)$ 
in Eq.~\eqref{condtens} are equal, and hence the 
 finite-$T$ conductance corrections are completely isotropic.  
Remarkably, all \textit{non-local}\ 
conductances, $G_{j\ne k}$ in Eq.~\eqref{condtens},
exhibit the same power-law temperature dependence and 
vanish at $T=0$, thereby
providing a highly characteristic signature to look for in experiments.
Indeed, the RAR scenario predicts $G_{j\ne k}=0$ at all $T$, while
the NFL manifold can be identified by a finite-$T$ non-local conductance
exhibiting power-law scaling.

\textit{Conclusions.---}In this work we have proposed a 
(challenging but realistic) device hosting a 
stable manifold of NFL states.  
By Josephson coupling a Majorana fermion system
to a superconductor, this suggests a novel route to a
first realization of this elusive behavior.
Future theoretical work should also study the 
full crossover from high to low temperatures, e.g.~using
numerical RG simulations \cite{nrg}.---
We thank A. Altland, A. Georges, P. Sodano, and A. Tsvelik for discussions,
and acknowledge financial support by the SFB TR12 and the SPP 1666 of the DFG.

\end{document}